\documentclass[10pt]{article}
\usepackage[OE]{express}

\newcommand{\p}{\partial}
\newcommand{\ep}{\varepsilon}

\newcommand{\om}{\omega}
\newcommand{\nn}{\nonumber}

\newcommand{\de}{\delta}

\begin{document}

\title{Multiple nonlinear resonances and frequency combs in bottle microresonators}

\author{I. Oreshnikov\authormark{1} and  D. V. Skryabin\authormark{1,2,*}}

\address{
\authormark{1}Department of Nanophotonics and Metamaterials, ITMO University, St. Petersburg 197101, Russia\\
\authormark{2}Department of Physics, University of Bath, Bath BA2 7AY, UK}
\email{\authormark{*}d.v.skryabin@bath.ac.uk} 


\begin{abstract}
We introduce the generalized Lugiato-Lefever equation describing nonlinear effects in the bottle microresonators. We demonstrate that the nonlinear modes of these resonators can form  multiple coexisting and overlapping nonlinear resonances and that their instabilities lead to the generation of the low repetition rate frequency combs.
\end{abstract}



\section{Introduction}
Physics and applications of the whispering gallery microresonators have been attracting a lot of interest over the past decade.
In particular, the frequency comb generation in the ring microresonators 
has revealed a plethora of complex nonlinear effects that included  
the formation of the so-called {\em soliton combs}, see, e.g., \cite{soliton1}. Microresonator combs and solitons are gradually finding their applications in the high precession spectroscopy \cite{spectr} and information processing \cite{koos}. The soliton combs in the microring resonators are the dissipative soliton pulses primarily relying on the balance between the Kerr nonlinearity and dispersion, that can be modeled using the Lugiato-Lefever (LL) equation \cite{soliton1}. The LL model was originally introduced to explain spatial patterns and localized states in the wide-aperture nonlinear resonators \cite{lug}. Apart from the microring resonators, the  bottle \cite{bot,pol1,snap,prl,dvo}, microbubble \cite{Farnesi2015,Lu2016}, spheroidal \cite{mat} and microsphere \cite{coen,suh}   resonators have also been developed for  sensing and frequency conversion applications. Four-wave mixing, nonlinear switching, Brillouin and Raman effects have been observed in the spheroidal \cite{mat}, bottle \cite{pol1,yang,Asano2016}, microbubble \cite{Farnesi2015,Lu2016} and microsphere \cite{coen,suh} resonators. 

In this work, we are proposing a generalization of the LL model to study nonlinear effects and frequency comb generation in the bottle microresonators. These resonators are made of silica or semiconductor strands/fibers (radius $r$) and operate  close to the cut-off frequency of a high order whispering gallery type mode. The side surface of the fiber is curved (bubbled) with the radius of curvature $R$, $R\gg r$, see Fig. 1(c). This curvature shapes the 'bottle' and provides the axial confinement, that transforms the slowly propagating waveguide mode into a discrete set of the resonator modes \cite{bot,snap,prl,dvo}. Dispersion near the waveguide cut-off is typically anomalous \cite{yulin,josab1,josab}, which together with the positive Kerr nonlinearity  creates conditions for the phase-matched cascaded four-wave mixing and soliton formation. Light can be coupled into the resonator using the evanescent tails of  a tapered fiber mode aligned perpendicular to the bottle axis. 

An advantage of the bottle resonators is that they can be integral parts of the surface nano-scale photonic circuits \cite{snap}. If used as the Kerr comb generators, the bottle resonators can cover a very wide range of the comb repetition rates, which equals the cavity free spectral range (FSR), from THz to MHz \cite{dvo}. This flexibility of the FSR design in microresonators has been first reported for microspheroidal resonators \cite{mat}, which have geometry and hence modal properties similar to the bottle ones. Importantly, bottle resonators  provide the GHz to MHz FSR values simultaneously with the on-chip integration option \cite{dvo}, while the ring resonators with comparable  FSRs are becoming too large for on-chip integration. Studying nonlinear effects in the low FSR devices is particularly interesting since the nonlinear shifts of the resonances start to compete and can exceed the FSR. Studies of these regimes  have started to emerge only recently in the context of the fiber loop resonators  \cite{wab}, where 100's of meters of fiber required to achieve the few MHz FSRs.
Below, we introduce the bottle resonator LL model and demonstrate that the modes of the  nonlinear bottle resonators can form multiple coexisting nonlinear resonances and their instabilities can lead to the generation of the low repetition rate frequency combs.

\begin{figure*}
\centering
\includegraphics[width=0.32\textwidth]{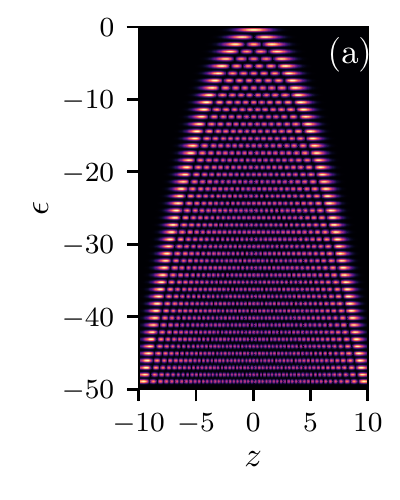}
\includegraphics[width=0.32\textwidth]{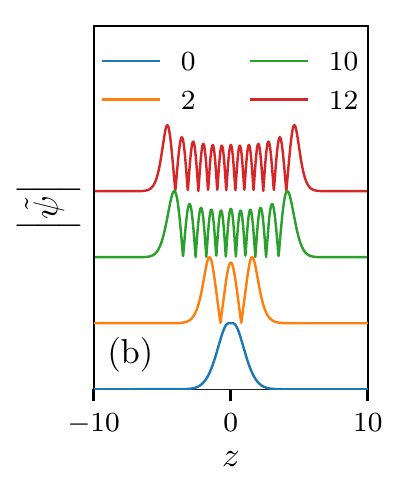}
\includegraphics[width=0.32\textwidth]{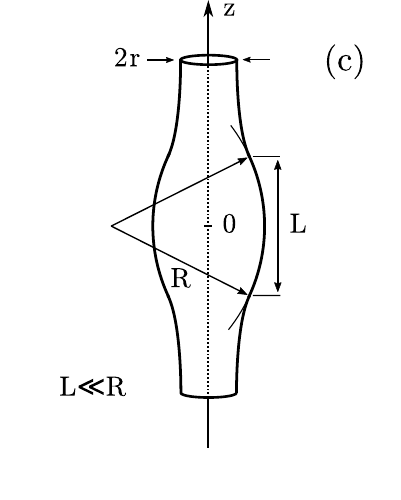}
\caption{(a) Density plot showing the spatial profiles vs the eigen-frequency $\epsilon$ of the $50$ modes in the linear bottle resonator.
(b) Spatial profiles, $|\tilde\psi(z)|$, of the upper branches of the nonlinear modes numerically computed using Eq. (\ref{eq3}) with $\p_t=0$: $z_p=0$, $P_0=1$, $\kappa=10^{-3}$; $\delta=0$ ($n=0,2$),  $\delta=-10$ ($n=10,12$). (c) Geometry of a bottle resonator.}
\label{f1}
\end{figure*}
\section{Lugiato-Lefever equation for bottle resonators}
The modal structure of  bottle resonators has been studied theoretically and experimentally \cite{snap,prl,dvo}. It was demonstrated that the linear Schr\"odinger equation with the parabolic potential and positive effective mass, corresponding to the anomalous group velocity dispersion, describes  the axial modal family belonging to one whispering gallery (azimuthal) resonance \cite{snap,prl}. The same equation, but including nonlinearity has been derived from the Maxwell equations in the context of the slow-light mode in optical fibers with longitudinal index variations \cite{josab1,josab}, which is exactly the technology used to fabricate bottle resonators \cite{bot,pol1,snap,prl}. For our purposes, we also need to account the pump and loss terms, which are well established in the optical resonator context, starting from the classical LL paper \cite{lug} and all the way to the recent  publications on the microresonator frequency combs, see, e.g., \cite{soliton1,spectr}. Thus the generalized LL equation for bottle microresonators is
\begin{equation}
i\p_T\Psi = - \frac{1}{2}d \p_Z^2\Psi 
     + \left(\om_0+\frac{f^2}{2d} U(Z)\right)\Psi-i\kappa_0\Psi-f|\Psi|^2\Psi-fP(Z)e^{-i\om_p T}.\label{eq1}
\end{equation}
Here $T$ is the physical time and $Z$ is the coordinate along the resonator axis. It is appropriate to term Eq. (\ref{eq1}) as the {\em generalized LL} equation, because it differs from the classical one \cite{soliton1,spectr,lug} by the potential term $U(Z)$  and the spatially varying pump $P(Z)=P_0e^{-(Z-Z_p)^2/w^2}$. Here $P_0$ is the dimensionless pump amplitude,  $Z_p$ is the position of the input nano-fiber running across the bottle and $w$ is the characteristic width of the evanescently coupled light \cite{snap,prl}.
$\omega_0$ is the reference frequency chosen close to the eigenfrequency of the ground states of the potential $U(Z)$, see below, and $\omega_p$ is the pump frequency.  $\kappa_0$ is the photon decay rate. $d>0$ is the group velocity dispersion coefficient \cite{yulin,josab1,josab}. $f$ is the FSR parameter,  which enters Eq. (\ref{eq1}) through an appropriate scaling of the dimensionless electric field envelope $\Psi$ and of the pump and potential terms. 

Linear spectrum of the pump and loss free resonator is found  assuming
$\Psi(Z,T)=\phi_n(Z)e^{-i\om_0T+i\ep_nT}$, which results in the eigenvalue problem
$-\ep_n\phi_n=-\frac{d}{2}\p_Z^2\phi_n+\frac{f^2}{2d}U(Z)\phi_n$.
Assuming $U(Z)=Z^2$, we recover the spectrum of the quantum harmonic oscillator: $\ep_n=-f (n+ 1/2)$. Thus $f$ is indeed the FSR, $\omega_0+f/2$ is the physical frequency of the fundamental cavity mode and $q=\sqrt{d/f}$ is the width of this mode.  One can assume as an estimate $f=60$~MHz, which physically  corresponds, e.g., to the silica cylinder with the  radius $r=300\mu$m and with the bottle curvature   $R=1$km: $f\simeq c/(3\pi)/\sqrt{rR}$, where $c=3\cdot 10^8$m/s~\cite{dvo}. $R\simeq 1$km has been demonstrated in  \cite{prl}. The number $N$ of modes in a bottle resonator of the length $L$ is estimated from $f^2 L^2/(2d)=f N$. Thus, $d=0.05Hz~m^2$ and $L=0.3$mm give the  fundamental mode width $q\simeq 30\mu$m and  $N\simeq 50$. Meaning of all geometrical parameters introduced above is clarified in Fig. 1(c). In what follows we restrict our simulations to the  $N=50$ case, however,  one can increase $N$ to thousands by increasing $L$ just to few mm's, $N\sim L^2$. $f|\Psi|^2$ represents the nonlinear shift of the resonance frequencies. We are primarily interested here in the case, when nonlinear effects are relatively large, so that $f|\Psi|^2$ is comparable to the FSR, $f$.
\begin{figure*}
  \centering
  \includegraphics[width=0.33\textwidth]{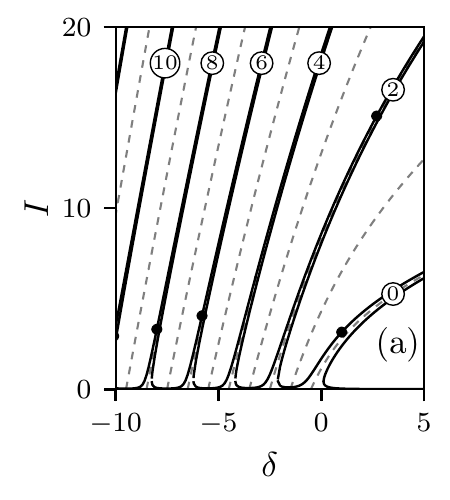}
  \includegraphics[width=0.33\textwidth]{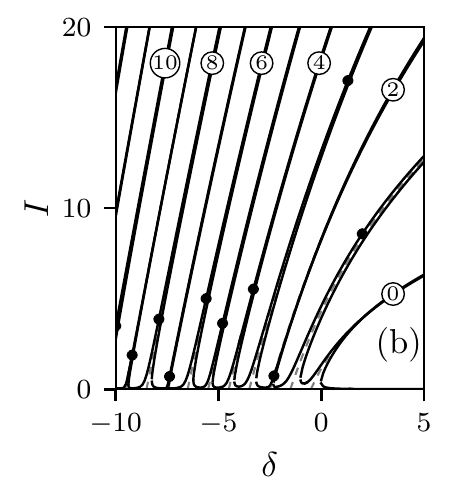}
  \includegraphics[width=0.33\textwidth]{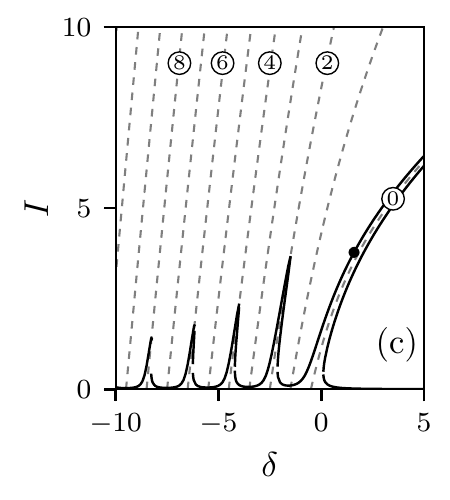}
  \includegraphics[width=0.33\textwidth]{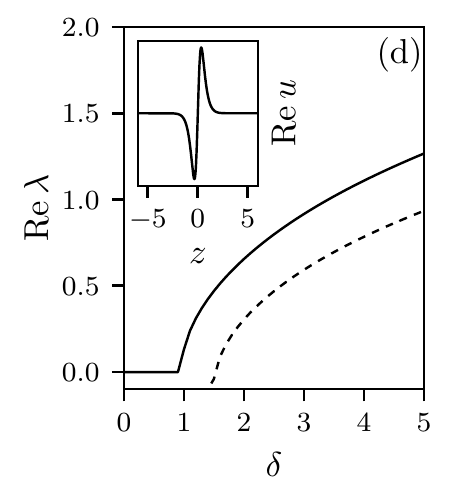}
  \caption{(a-c) Structure of the multiple nonlinear resonances for the different pump positions and the resonator losses. Plots show $I=\int_{-\infty}^{\infty}|\tilde\psi(z)|^2dz$ vs $\de$. (a) $z_p=0$ (maximum of the ground state mode), $\kappa=10^{-3}$, (b) $z_p=0.79$ (maximum of the $n=1$ mode), $\kappa=10^{-3}$, (c) $\kappa=10^{-1}$, $z_p=0$.  The bold points show the onsets of the instabilities of the upper branches of the nonlinear resonances, which are stable below these points. The encircled numbers correspond to the modal index $n$. (d) The instability growth rates $\mathrm{Re}\, \lambda$ vs $\delta$ of the upper branch of the ground state resonance. Parameters as in (a) - solid line, and (c) - dotted line. The inset shows $\mathrm{Re}\, u(z)$ of the unstable perturbation.}
 \label{f2}
\end{figure*}

For our numerical studies we have dimensionalized Eq. (\ref{eq1}) to the form
\begin{equation}
i\p_t\psi=-(1/2)\p_z^2\psi+\left(\delta+U(z)/2\right) \psi-i\kappa\psi-|\psi|^2\psi-P(z),\label{eq3}
\end{equation} 
where $\psi=\Psi e^{i\omega_pT}$,  scaled time is $t=fT$,  distance is  $z=Z/q$,  $\kappa=\kappa_0/f$ is the normalized loss rate and $\delta=(\omega_0-\omega_p)/f$ is the normalized detuning.
$U(z)=z^2$ for $z<l$ and $U(z)=l^2$ for $z>l$, where  $l=L/q$ is the scaled resonator length, $l^2/2=N$ and $N=50$.
Dimensionless pump width $w/q$ is $0.2$ in the numerical examples shown. The density plot in Fig. \ref{f1}(a) shows  $|\psi(z)|$ for the linear modes, calculated from $-2\epsilon~\psi=-\p_z^2\psi+U(z) \psi$, vs their eigen-frequencies, $\epsilon\simeq -(n+1/2)$.
\section{Multistability, instabilities and comb generation}
For $\kappa=P=0$, the nonlinear modes split from the linear spectrum at points $\delta=-(n+1/2)$ and  their amplitudes increase with $\de$ through the positive Kerr nonlinearity, see dashed lines in Figs. \ref{f2}(a)-\ref{f2}(c).
As soon as the pump and loss are introduced, the dashed lines split into pairs. If $P(z)=P(-z)$, as for $z_p=0$, then the odd modes are not excited, see Figs. \ref{f2}(a) and \ref{f2}(c). Otherwise, the even and odd modes show up together, see Fig. \ref{f2}(b). In the  $\kappa=0$ limit, the nonlinear modes extend to $\de\to+\infty$, so that for any $\delta>0$ we have $N$ co-existing solutions: {\em multistability}. As $\kappa$ increases the intervals of $\delta$, where the tilted nonlinear resonances  exist, become narrower and eventually shrink below the FSR, at what point  the {\em multistability} is replaced by the more usual {\em bistability}, cf. Figs. \ref{f2}(a,b) and \ref{f2}(c). Data for  Figs. \ref{f2}(a)-\ref{f2}(c) were obtained by numerically solving Eq. (\ref{eq3}) assuming $\psi(z,t)=\tilde\psi(z)$.  Some examples of the nonlinear mode profiles $\tilde\psi$  are shown in Fig. \ref{f1}(b).

In order to study stability of the nonlinear modes $\tilde\psi$ with respect to  noise, we have linearized Eq. (\ref{eq3})  using the substitution $\psi(z,t)=\tilde\psi(z)+u(z)e^{\lambda t}+v^*(z)e^{\lambda^* t}$ with $|\tilde\psi|\gg |u|,|v|$ and solved the resulting eigenvalue problem,
\begin{eqnarray}
&& \lambda u=-i\left(\delta+U(z)/2\right) u+(i/2)\p^2_zu-\kappa u+2i|\tilde\psi|^2u+i\tilde\psi^2v, \label{eq4}\\ \nn &&
\lambda v=i\left(\delta+U(z)/2\right) v-(i/2)\p^2_zv-\kappa v-2i|\tilde\psi|^2v-i\tilde\psi^{*2}u,
\end{eqnarray}
numerically.
\begin{figure*}
\centering
\includegraphics[width=\textwidth]{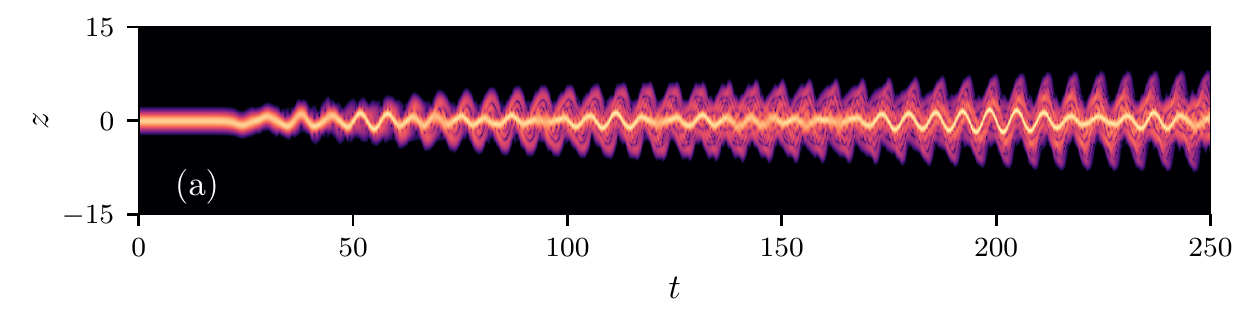}
\includegraphics[width=0.3\textwidth]{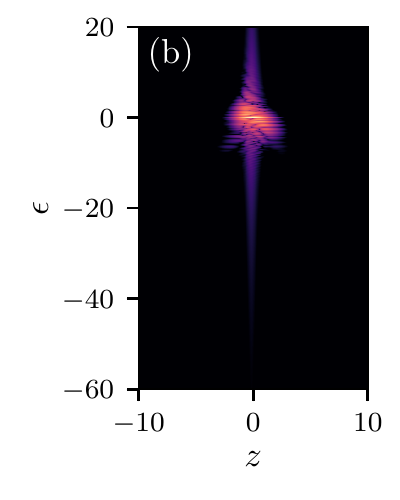}
\includegraphics[width=0.3\textwidth]{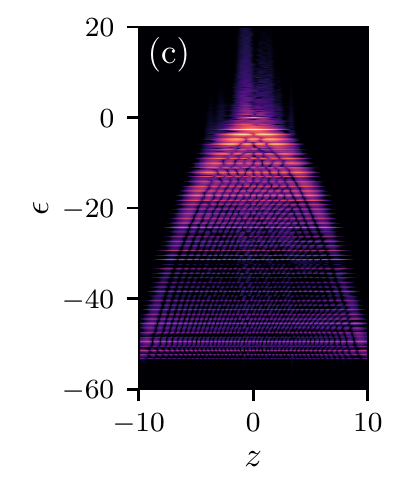}
\includegraphics[width=0.3\textwidth]{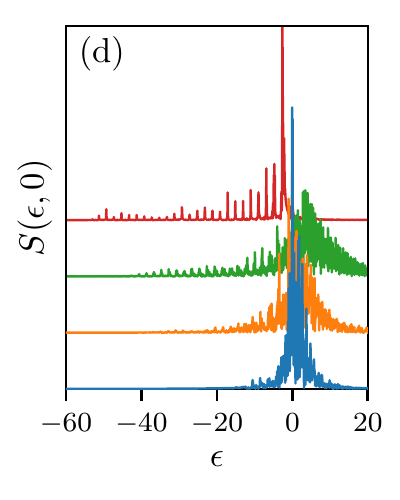}
\caption{(a) Instability development of the ground state mode. Parameters as in Fig. \ref{f1}(a), $\delta=3$. Spectra of the field across the resonator at the initial (b) and the advanced  (c) stages of the comb generation calculated as $S(\ep,z)=|\int_{t_0}^{t_0+\tau}\psi(z,t)e^{-i\epsilon t}dt|$: $t_0=0$ in (b),
  $t_0=1450$ in (c) and $\tau=50$.   (d) The spectra at $z=0$ and $\tau=50$:  $t_0=0$ (blue); $t_0=400$ (orange); $t_0=900$ (green); $t_0=1400$ (red). $\epsilon$ in (b)-(d) corresponds to the physical frequency $\omega_p-f\ep$.}
  \label{f3}
\end{figure*}
\begin{figure*}
\centering
\includegraphics[width=\textwidth]{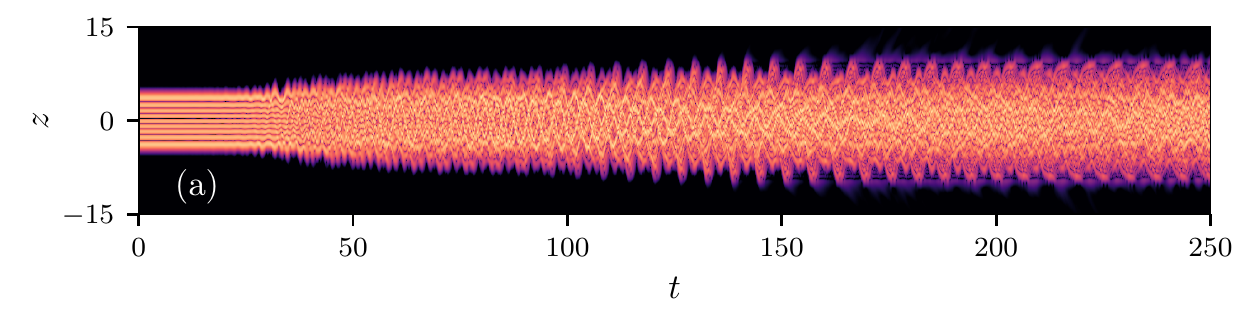}
\includegraphics[width=0.3\textwidth]{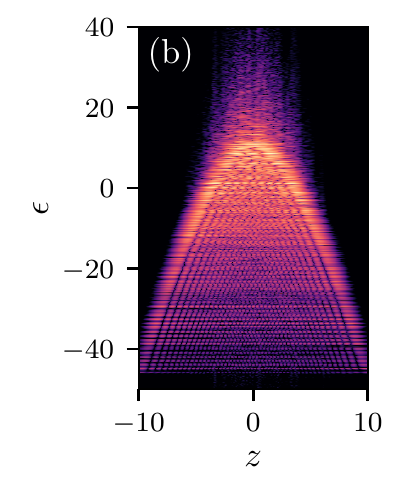}
\includegraphics[width=0.3\textwidth]{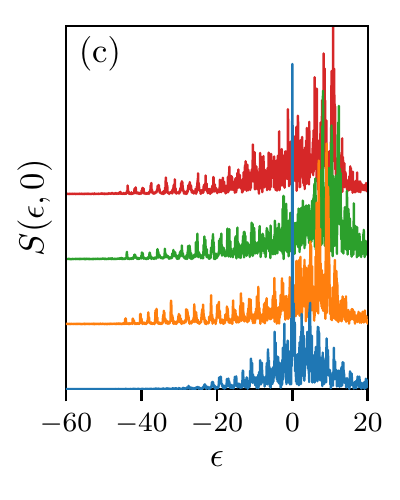}  
\caption{
(a) Instability development of the $n=8$ mode. Parameters as in Fig. \ref{f1}(a), $\delta=-5$. (b) Spectrum $S(\ep,z)$, see Fig. \ref{f3},  at the advanced   stage of the comb generation: $t_0=1450$, $\tau=50$.  (c) The spectra at $z=0$. The colors and the time intervals same as in Fig. \ref{f3}(d).}
\label{f4}
\end{figure*}
In the LL equation without the potential term and with the homogeneous pump, the upper branch of the homogeneous solution is always unstable in the case of focusing nonlinearity \cite{lug}. The harmonic potential spatially confines the modes,  and therefore the instability range shrinks, which is particularly pronounced for the modes with the low to moderate $n$'s, see Fig. \ref{f2}. In particular, we have found the coexistence of the stable upper branches of the $n=0$ and $n=2$ resonances, Fig. \ref{f2}(a), and of the 
$n=0,1$ and $3$ ones, see Fig. \ref{f2}(b). Of course, the lowest amplitude solution is also stable at the same time. Thus  the true multistability is realized in our system, which is a relative rare situation in nonlinear optical devices. Increasing the losses leads to broadening of the resonances and either reduction or complete suppression of all the instabilities,  Fig. \ref{f2}(c).

The inset in Fig. \ref{f2}(d) shows the eigensolution of Eqs. (\ref{eq4}) driving the instability of the ground state $n=0$ and the figure itself shows the corresponding growth rate as a function of $\delta$. Thus the ground state becomes  unstable above some critical detuning with respect to the perturbations that are shaped like $n=1$ mode. The noise driven excitation of this instability leads to the development of the periodic in time and space oscillations of the localized wavepacket in the harmonic potential, see Fig. \ref{f3}(a). These oscillations lose their regularity with time, while more of the higher order modes are getting excited, so that the spectral content of the field is broadening and the frequency comb is generated. Figs. \ref{f3}(b) and \ref{f3}(c) show the spectra of the intracavity field calculated at every point inside the resonator at the initial stage of the evolution and when the frequency comb has already fully developed. Fig. \ref{f3}(d) shows how the  spectrum at  $z=0$ changes with time and  acquires the comb structure.  Since, practically, the signal is going to be collected at a point, this is the type of spectrum which is expected to be seen in the experiments. By looking at the space-time evolution of the field in Fig. \ref{f3}(a), one can say that this comb corresponds to the quasi-soliton pulse oscillating in the harmonic potential and immersed into the sea of the weakly nonlinear modes represented by the discrete part of the spectrum for $\epsilon<-1/2$.
The  dynamics generally gets more complex as the pump frequency is tuned into a resonance with the higher order modes. Fig. \ref{f4} shows the space-time and spectral evolution observed when we initialized Eq. (\ref{eq3}) with the unstable $n=8$ mode. In this case,  several localized wavepackets emerge and interact  in the potential between themselves and with the extended nonlinear modes. This produces the combs, see Figs. 
\ref{f4}(b) and \ref{f4}(c), which are both broader and more intense than the ones in Fig. \ref{f3}. 
\section{Summary}
We have introduced a generalization of the LL model applicable to the bottle resonators and demonstrated  multistability and  generation of the low repetition rate combs in these devices.  
\section*{Funding}
The Leverhulme Trust (RPG-2015-456); H2020 (691011, Soliring); ITMO University (Grant 074-U01); RFBR (17-02-00081); RSCF (17-12-01413).
\end{document}